\def \SAIT #1 #2 {{\em Mem.\ Soc.\ Astron.\ It.\/} {\bf #1}, #2}
\def \MESS #1 #2 {{\em The Messenger\/} {\bf #1}, #2}
\def \ASTRNACH #1 #2 {{\em Astron. Nach.\/} {\bf #1}, #2}
\def \AAP #1 #2 {{\em Astron. Astrophys.\/} {\bf #1}, #2}
\def \AAL #1 #2 {{\em Astron. Astrophys. Lett.\/} {\bf #1}, L#2}
\def \AAR #1 #2 {{\em Astron. Astrophys. Rev.\/} {\bf #1}, #2}
\def \AAS #1 #2 {{\em Astron. Astrophys. Suppl. Ser.\/} {\bf #1}, #2}
\def \AJ #1 #2 {{\em Astron. J.\/} {\bf #1}, #2}
\def \ANNREV #1 #2 {{\em Ann. Rev. Astron. Astrophys.\/} {\bf #1}, #2}
\def \APJ #1 #2 {{\em Astrophys. J.\/} {\bf #1}, #2}
\def \APJL #1 #2 {{\em Astrophys. J. Lett.\/} {\bf #1}, L#2}
\def \APJS #1 #2 {{\em Astrophys. J. Suppl.\/} {\bf #1}, #2}
\def \APSS #1 #2 {{\em Astrophys. Space Sci.\/} {\bf #1}, #2}
\def \ASR #1 #2 {{\em Adv. Space Res.\/} {\bf #1}, #2}
\def \BAIC #1 #2 {{\em Bull. Astron. Inst. Czechosl.\/} {\bf #1}, #2}
\def \JSQRT #1 #2 {{\em J. Quant. Spectrosc. Radiat. Transfer\/} {\bf #1}, #2}
\def \MN #1 #2 {{\em Mon. Not. R. Astr. Soc.\/} {\bf #1}, #2}
\def \MEM #1 #2 {{\em Mem. R. Astr. Soc.\/} {\bf #1}, #2}
\def \PLR #1 #2 {{\em Phys. Lett. Rev.\/} {\bf #1}, #2}
\def \PASJ #1 #2 {{\em Publ. Astron. Soc. Japan\/} {\bf #1}, #2}
\def \PASP #1 #2 {{\em Publ. Astr. Soc. Pacific\/} {\bf #1}, #2}
\def \NAT #1 #2 {{\em Nature\/} {\bf #1}, #2}

\documentstyle{memsait}
\input epsf.sty
\begin{opening}
\title{The (anti)correlation of the sub-mm and X-ray background sources} 
\author{P. Severgnini$^{1}$}
\institute{Universit\`a di Firenze, Italy}
\date{} 
\end{opening}

\begin{document}

\oddpagefooter{}{}{} 
\evenpagefooter{}{}{} 
\ 
\bigskip

\begin{abstract}
The connection between the sub-mm and the hard X-ray backgrounds is studied
by comparing data at 2--10 keV and at 850$\mu$m for a sample of 34
sources at fluxes (or limiting fluxes) which resolve most of the
background in the two bands. These data were obtained with
new SCUBA observations and by correlating data sets available from the
literature. None of the 11 hard X-ray (2--10 keV) sources has a
counterpart at 850$\mu$m, with the exception of a faint
Chandra source, which
is a candidate type 2 QSO at high redshift.
These data indicate that 2--10 keV sources brighter than
$10^{-15} erg~s^{-1}cm^{-2}$, which make at least 75\% of the background
in this band, contribute for less than 7\% to the submillimetric
background. Out of the 24 SCUBA sources 23 are undetected by Chandra.
These data indicate that most of these SCUBA sources must be powered
either by starburst activity, or by an AGN which is obscured by a column
$\rm N_H > 10^{25} cm^{-2}$, with a reflection efficiency in the hard
X-rays significantly lower than 1\% in most cases.
\end{abstract}

\vskip0.5truecm

\footnotetext[1]{On behalf of a collaboration including
R. Maiolino, M. Salvati, D. Axon, A. Cimatti, F. Fiore, R. Gilli,
F. La Franca, A. Marconi, G. Matt, G. Risaliti, and C. Vignali.}

Following the original suggestion of Setti \& Woltjer (1989), the
hard X-ray background (2--100 keV) is commonly explained with the superposed
emission of a large population of highly obscured, type 2 AGNs.
However, obscured AGNs have also been proposed as the powerhouse of
a fraction of the SCUBA sources which make most of the background at 850$\mu$m,
thus providing a link between the two spectral windows. In particular,
some models predict that obscured AGNs could contribute a substantial
fraction (20--50\%) of the submillimeter background (Almaini et al. 1999).

With the aim of investigating the relation between the sub-mm and the
2--10 keV hard X-ray backgrounds, we started an observing program
with SCUBA of a subsample of the sources detected by BeppoSAX in the
HELLAS survey. This survey resolved about 30\% of the 5--10 keV background
into discrete sources at a flux limit of $\rm \sim 5\times 10^{-14} erg
~s^{-1}cm^{-2}$. None of the four hard X-ray sources observed so far was
detected at 850$\mu$m at a 2$\sigma$ limiting flux of about 3 mJy.
In order to put these results into perspective, we follow Fabian et al.
(2000) and compute a submillimeter--to--X-ray index $\alpha_{SX}$
($\rm F_{\nu}\propto \nu ^{-\alpha}$), using the observer-frame flux
densities at 850$\mu$m and 5 keV. In Fig.~1, which gives $\alpha_{SX}$
as a function of redshift, our results are represented by filled squares.

We have also cross-correlated the deep hard X-ray survey of the SSA13 field
($\rm F_{2-10keV}>2.5\times 10^{-15} erg~s^{-1}cm^{-2}$),
carried out by Mushotzky et al. (2000), with the deep (small-area) and
shallow (wide-area) SCUBA maps of the same field
obtained by Barger et al. (1999) at 850$\mu$m. Only one out
 of the three Chandra
sources in the deep SCUBA map is detected at 850$\mu$m, which is
indicated with a filled circle in Fig.~1. The optical, X-ray and sub-mm
properties of this object suggest that this is a candidate type 2 QSO,
heavily absorbed and hosted in a young galaxy at redshift between 1.9
and 2.7. The other two Chandra sources in this area, undetected at 850$\mu$m,
are marked as upper limits with filled circles in Fig.~1. Eight out of the
nine SCUBA sources in the SSA13 field are not detected by Chandra; since
their redshift is not known the spread of the lower limits on $\alpha _{SX}$
is indicated with two horizontal shaded areas in Fig.~1.

The open squares in Fig.~1 indicate the limits on $\alpha _{SX}$
for the sources in the lensing clusters A1835 and A2390
obtained by Fabian et al. (2000) by cross-correlating Chandra observations
and SCUBA maps. Finally, the open circles are the limits on
$\alpha _{SX}$ for the sources in the Hubble Deep Field North obtained
by Hornschemeier et al. (2000), or derived by us, with a deep Chandra
observation ($\rm F_{2-8keV}>9\times 10^{-16} erg~s^{-1}cm^{-2}$).

In Fig.~1 we also show the values of $\alpha _{SX}$ as a function of
redshift for various classes of objects. NGC4593 and the two PG QSOs are
assumed to be representative of type 1 Seyferts and of radio quiet QSOs.
Four QSOs at high-z detected by SCUBA are shown with stars.
Circinus is shown as
representative of heavily obscured Sy2s ($\rm N_H = 5\times 10^{24} cm^{-2}$),
but we also show the expected $\alpha _{SX}$ in the case that the absorbing
column density were an order of magnitude lower than observed
(namely $\rm N_H = 5\times 10^{23} cm^{-2}$). NGC1068 is the case of a
completely Compton thick AGN ($\rm N_H > 10^{25} cm^{-2}$) whose hard
X-ray flux is reflection dominated up to 100 keV. As a starburst
template we show the case of Arp220.

The thick upper horizontal segment in Fig.~1 gives the $\alpha _{SX}$ of the
cosmic background, under the assumption that most of the flux in both spectral
windows comes from redshifts between 1 and 2 (in analogy with the soft
X-ray background). The location of
the $\alpha _{SX}$ of the various classes of sources with
respect to this value constrains their contribution
to the sub-mm and to the hard X-ray background.
In particular,
the lower thick horizontal segment gives the $\alpha _{SX}$ of sources
which would contribute 100\% of the 2-10 keV background and only 10\% of the
sub-mm background.

Out of the 11 hard X-ray sources only one is detected at 850$\mu$m.
The upper limits on $\alpha _{SX}$ for the hard X-ray sources are much
lower than the background requirements. In particular, we estimate that,
under conservative assumptions, the 2--10 keV sources brighter than
$\sim  10^{-15} erg~s^{-1}cm^{-2}$, which resolve at least 75\% of the
background in this band (Mushotzky et al. 2000), cannot contribute for
more than 7\% to the sub-mm background. A more detailed discussion
is given in Severgnini et al. (2000).

None of the 24 SCUBA sources, but the candidate type 2 QSO in the SSA13
field, is detected in the 2--10 keV band down to a
limiting flux of $\rm F_{2-10keV} \sim 1-2\times 10^{-15}
erg~s^{-1}cm^{-2}$. As shown in Fig.~1, their lower limits on $\alpha _{SX}$
occupy the upper part of the plot and are presumably akin to the
starburst template given by Arp220. Most of the SCUBA sources with known
redshift are even inconsistent with the reflection-dominated template
given by the nucleus of NGC1068. These sources are probably dominated by
starburst activity. A significant contribution from an obscured AGN might
be present if the latter is completely Compton thick
($\rm N_H > 10^{25} cm^{-2}$) and if the reflection efficiency is
significantly lower than estimated for NGC 1068 ($\sim$1\%).


\begin{figure}
\epsfxsize=11cm 
\hspace{5.0cm}\epsfbox{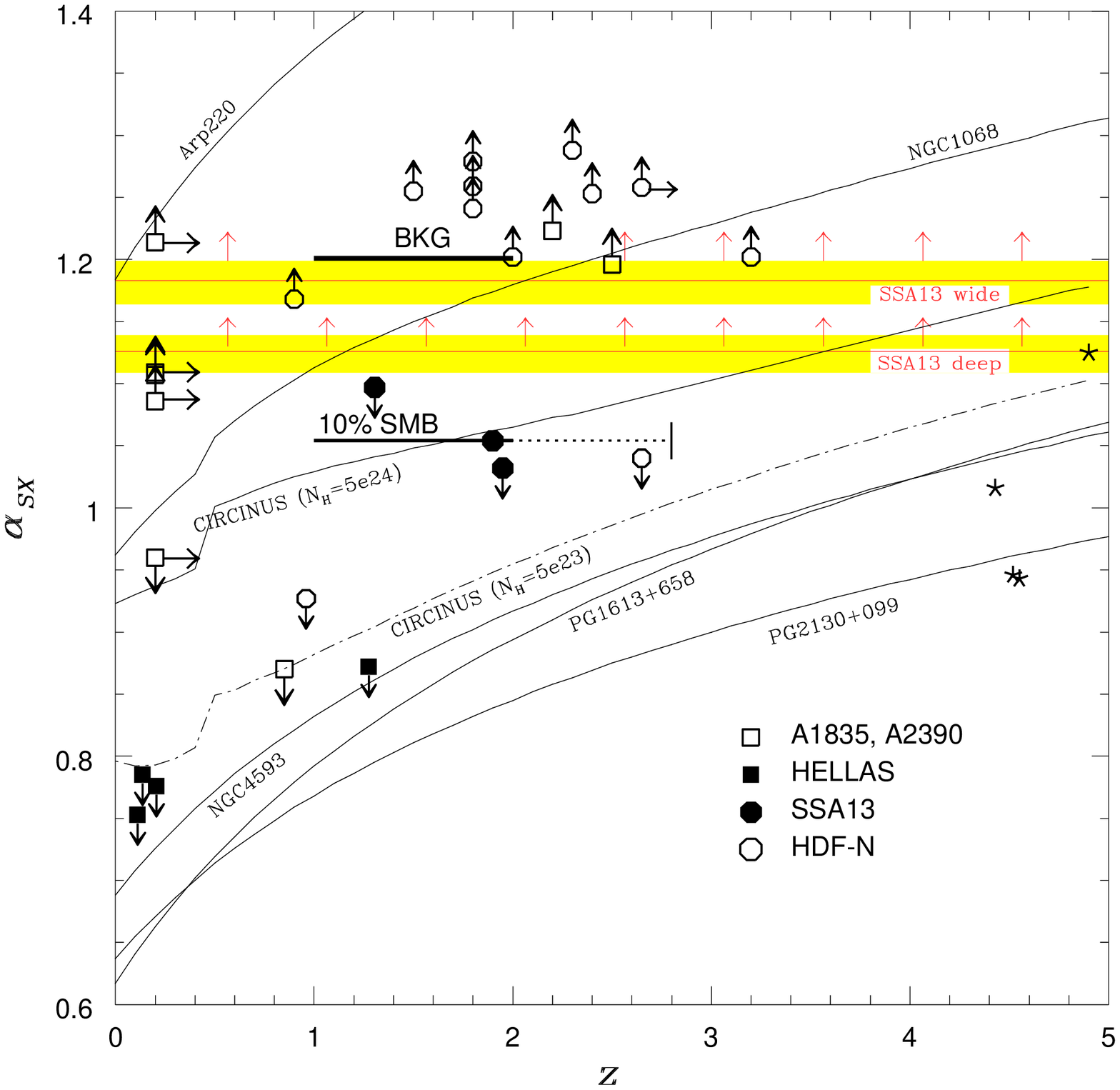} 
\caption[h]{}
\end{figure}


\acknowledgements
This work was partially supported by the Italian Space Agency (ASI)
through the grant ARS-99-75 and by the Italian Ministry for
University and Research (MURST) through the grant Cofin-98-02-32.


\end{document}